\documentclass[3p,times]{elsarticle}

\usepackage{ecrc}

\volume{00}

\firstpage{1}

\journalname{Nuclear Physics A}

\runauth{Hongxi Xing, Yun Guo, Enke Wang and Xin-Nian Wang}

\jid{nupha}

\usepackage{amssymb}

\usepackage[figuresright]{rotating}

\begin{document}

\begin{frontmatter}



\dochead{}

\title{Parton energy loss in cold nuclei}


\author{Hongxi Xing$^{a}$, Yun Guo$^{c}$, Enke Wang$^{a}$ and Xin-Nian Wang$^{a,b}$}

\address{$^a$ Institute of Particle Physics, Central China Normal University, Wuhan 430079, China\\
$^b$Nuclear Science Division, MS 70R0319, Lawrence Berkeley National Laboratory, Berkeley, CA 94720\\
$^c$Department of Physics, Guangxi Normal University, Guilin 541004, China}

\begin{abstract}
Within the generalized high-twist factorization formalism, we express
the contribution from multiple parton scattering and induced gluon
radiation to the DY dilepton spectra in terms of nuclear modified
effective beam quark distribution function. We show that beam quark
energy loss is characterized by jet transport parameter $\hat{q}$,
which is related to the local gluon density of the medium. Using the
value of $\hat{q}$ determined from the deeply inelastic scattering
(DIS) data, we evaluate the nuclear modification factor in the Drell-Yan
process in p+A collisions. Effects of parton energy loss in the DY
spectra are found negligible in the Fermilab experimental data at
$E_{\rm lab}=800$ GeV relative to parton shadowing while the predicted
suppression of the DY spectra are significant at $E_{\rm lab}=120$ GeV.
\end{abstract}


\end{frontmatter}


\section{Introduction}
\label{Introduction}
Jet quenching is one of the most  important discoveries in heavy-ion collisions at RHIC. It can provide crucial information of the properties of the hot and dense matter \cite{Gyulassy:1990ye}, quark gluon plasma (QGP), which is expected to be formed under extreme conditions. However, in A+A collisions, both final-state QGP effects, such as jet quenching, and initial-state effects in cold nuclear matter can modify the experimental observables. In pursuit of the real signals from QGP, one has to study the baseline initial conditions in A+A collisions, which include transverse momentum broadening, parton shadowing and parton energy loss in cold nuclei.

In this study, based on generalized high-twist factorization formalism \cite{Luo:1992fz}, we discuss the initial-state effects of parton energy loss in the Drell-Yan process in p+A collisions due to multiple scattering inside the nuclear medium. The calculation is similar to final-state multiple scattering in deeply inelastic scattering (DIS) off large nuclei, where the medium modification to the quark fragmentation function caused by multiple scattering and induced gluon bremsstrahlung is effectively written as \cite{Guo:2000nz},
\begin{eqnarray}
\label{Eq mff}
\tilde{D}_q^h(z_h,\mu^2)&=&D_q^h(z_h,\mu^2)+\frac{\alpha_s}{2\pi}\int_0^{\mu^2}\frac{dl_T^2}{l_T^2}
\int_{z_h}^{1}\frac{dz}{z}\left[\Delta\gamma_{q\rightarrow qg}(z,l_T^2)D_q^h(z_h/z)+\Delta\gamma_{q\rightarrow gq}(z,l_T^2)D_g^h(z_h/z)\right]\, ,
\end{eqnarray}
where the medium modified splitting functions $\Delta\gamma_{a\rightarrow bc}(z,l_T^2)$ depend on the properties of the medium via the jet transport parameter which is related to the local gluon density,
\begin{eqnarray}
\hat{q}_F(\xi_N)=\frac{4\pi^2\alpha_sC_F}{N_c^2-1}\rho_A(\xi_N)[xG(x)]_{x\approx 0},
 \end{eqnarray}
where $\rho_A$ is the nuclear density distribution and $G(x)$ is the gluon distribution function in a nucleon. Based on the modified fragmentation function from single induced gluon emission, one can derive a medium modified DGLAP evolution equation, and determine the value of $\hat{q}$ from comparisons to the HERMES experimental data \cite{Airapetian:2007vu}.  A recent study finds a range of the jet transport parameter $\hat{q}_0=0.016-0.032$GeV$^2$/fm in cold nuclei by fitting the HERMES data on $\pi^{\pm}$ and $K^{\pm}$ spectra in SIDIS \cite{Wang:2009qb} and this value is consistent with the transverse momentum broadening of the Drell-Yan dilepton production in p+A collisions \cite{Guo:1998rd}. In addition, a recent study shows that one can also obtain the value of $\hat{q}$ by studying the nuclear enhancement of transverse momentum imbalance for two back-to-back particle production in e+A and p+A collisions \cite{Kang:2011bp}.

\section{Parton energy loss in cold nuclei}
\label{}
In the Drell-Yan process in p+A collisions, the energetic quark coming from the beam hadron will undergo multiple scattering with the target remnant before it annihilates with an antiquark from the target nucleus to produce dileptons. The contribution from multiple scatterings in nuclear medium can be calculated by the generalized high-twist factorization formalism in pQCD \cite{Luo:1992fz}. In such a factorization formalism, one has to expand the hard partonic part of the hard interaction in the intrinsic parton transverse momentum. In the expansion, the first term gives rise to the normal collinear factorized pQCD results, known as leading twist contributions. The high-order terms in the Taylor expansion are known as higher-twist contributions which can be expressed as the convolution of hard partonic parts and high-twist matrix elements inside the target nuclei. In general, the twist-4 contribution can be expressed as
\begin{eqnarray}
\label{Eq twist4 c-section} \nonumber \frac{d\sigma_{qA \rightarrow
\gamma^*}^D}{d^4q}&=&\frac{1}{2\xi s}
\int\frac{dy^-}{2\pi}\frac{dy_1^-}{2\pi}\frac{dy_2^-}{2\pi}
\frac{1}{2}\langle A|F^+_\alpha(y_2^-)\bar{\psi}_q(0)\gamma^+
\psi_q(y^-)F^{+\alpha}(y_1^-)|A\rangle\\
&\times&(-\frac{1}{2}g^{\alpha\beta})\left[\frac{1}{2}\frac{\partial^2}{\partial{k_T^\alpha}
\partial{k_T^\beta}}\overline{H}(y^-,y_1^-,y_2^-,k_T,p,q)\right]_{k_T=0},
\end{eqnarray}
where $\xi$ is the momentum fraction carried by the initial parton in the beam hadron and $k_T$ the intrinsic transverse momentum of the initial gluon in the nuclear target.
In our study, we consider two distinctive double scattering processes as shown in Figs.4 and 5 in \cite{Xing:2011fb} . One is the annihilation-like process with the additional scattering between the beam quark and a gluon from the nucleus; the other is the Compton-like process, where the additional scattering is between two gluons coming from the beam hadron and target nucleus, respectively. Besides these two double-scattering processes, one also needs to consider the interferences between single and triple scatterings. Contributions from the first term in the collinear expansion from all the processes mentioned above give the eikonal contributions to the single scattering to ensure the gauge invariance of the leading-twist results. The second derivative of the hard partonic part with respect to $k_T$ gives rise to the twist-4 contributions. However, the contributions from the interferences between single and triple scatterings are power suppressed ($q_T^2/Q^2$) as compared to the leading-twist-4 contributions from double scattering. Within the generalized high-twist factorization formalism as shown in Eq. \ref{Eq twist4 c-section}, one can calculate the final leading logarithmic contributions coming from the annihilation-like and Compton-like processes. Summing up all the leading logarithmic contributions from single and double scattering, we can express the contributions from multiple scattering and induced gluon radiation to the DY dilepton spectra in terms of nuclear modified effective beam quark distribution function \cite{Xing:2011fb} ,
\begin{eqnarray}
\label{Eq mpdf} \nonumber
\tilde{f}_{q/h}(x',\mu^2,A)&=&f_{q/h}(x',\mu^2)+\frac{\alpha_s}{2\pi}\int_0^{\mu^2}\frac{dq_T^2}{q_T^2}
\int_{x'}^{1}\frac{d\xi}{\xi}\left[f_{q/h}(\xi)\Delta\gamma_{q\rightarrow qg}(x'/\xi,q_T^2)\right.\\
&+&\left.f_{g/h}(\xi)\Delta\gamma_{g\rightarrow
q\bar{q}}(x'/\xi,q_T^2)\right]\, ,
\end{eqnarray}
with $x^{\prime}$ is the momentum fraction carried by the beam quark in the projectile hadron. The modified beam quark distribution function takes a form very similar to the vacuum bremsstrahlung corrections that lead to the  evolution equations in pQCD for parton distribution functions. In Eq. (\ref{Eq mpdf}), $f_{q/h}(x',\mu^2)$ is the renormalized twist-two beam quark distribution. The nuclear-dependence of the medium-modified beam quark distribution function is implicit through the medium-modified splitting functions $\Delta\gamma_{q\rightarrow qg}$ and $\Delta\gamma_{g\rightarrow
q\bar{q}}$  which are defined in Eq. (32) and (37) in \cite{Xing:2011fb}.

The medium-modified splitting functions are proportional to twist-4 matrix element $T_{g\bar q}$ which can be factorized in terms of quark and gluon density distribution inside the target nucleus under the assumption that nucleon correlation inside nuclei is negligible. The gluon density in turn can be related to quark transport parameter $\hat{q}$ in the nuclear medium, therefore,
\begin{eqnarray}
T_{g\bar q}(x,x_t)&\approx&\frac{6\hat{q}_0 f_{\bar
q/N}(x)}{\pi\alpha_s \rho_A(0,\vec{0}_{\perp})}\int
d^2{b}\int_{-\infty}^{\infty}dy^-\int_{-\infty}^{y^-}
dy_1^- \rho_A(y_1^-,\vec{b})\rho_A(y^-,\vec{b}){\rm sin}^2(x_tp^+y_1^-/2),
\end{eqnarray}
where we have used the assumption that the jet transport parameter along the quark jet trajectory is proportional to the nuclear density $\hat{q}(y,b)=\hat{q}_0\rho_A(y,b)/\rho_A(0,0)$, and we will employ the Woods-Saxon nuclear geometry in our numerical calculation. In the definition of  $T_{g\bar q}$, the trigonometric function comes from the LPM interference which plays an important role in the small $q_T^2$ region and leads to the quadratic nuclear size dependence of the parton energy loss.

\begin{figure}[h]
\begin{center}
\includegraphics[width=14.0cm]{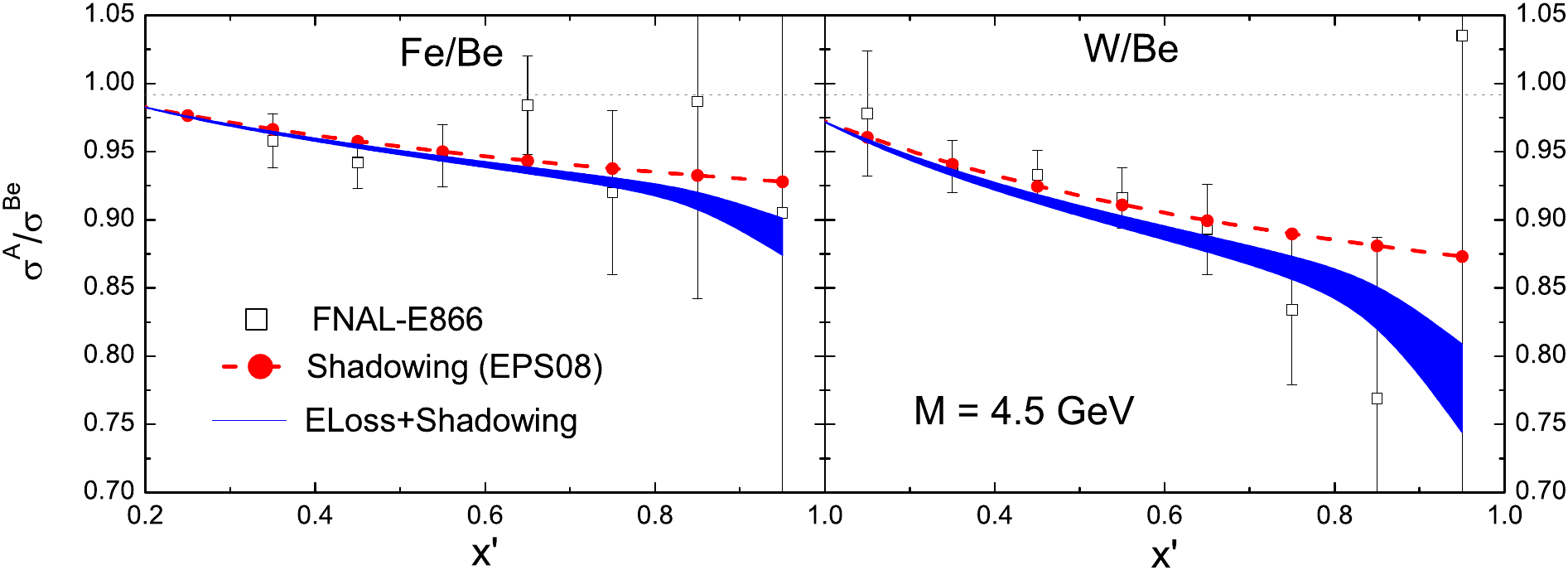}
\centerline{\parbox{15cm}{\caption{\label{Fig-e866}\small Ratios of
the DY cross section per nucleon in $p+A$ collisions versus $x'$at $E_{lab}=800$ GeV. The shaded bands
correspond to $\hat{q}_0=0.024\pm0.008~\rm GeV^2/fm$
\cite{Wang:2009qb}. The experimental data are from the Fermilab
experiment E866 \cite{Vasilev:1999fa}. }}}
\end{center}
\end{figure}

Shown in Fig. \ref{Fig-e866} are the calculated nuclear modification factors at leading order,
\begin{eqnarray}
\label{ratio}
\frac{B \sigma^A}{A \sigma^B}=\frac{B d \sigma_{pA\rightarrow
l^+l^-}/dQ^2dx'}{ A d\sigma_{pB\rightarrow l^+l^-}/dQ^2dx'}
=\frac{\sum_q \int dx
f_{\bar{q}/A}(x,\mu^2)\tilde{f}_{q/p}(x',\mu^2,A)H_0(x,p,q)}{A \sum_q
\int dx f_{\bar{q}/N}(x,\mu^2) f_{q/p}(x',\mu^2)H_0(x,p,q)},
\end{eqnarray}
as a function of $x^{\prime}$ for the Drell-Yan dilepton production as compared to the Fermilab E866 experimental data \cite{Vasilev:1999fa}.

For fixed invariant mass $M$, the fractional momentum $x=M^{2}/x^{\prime}s$ carried by the target partons becomes smaller for large beam quark fractional momentum $x^{\prime}$ in the kinematic region of the E866 experiment, therefore strong nuclear shadowing of the quark distribution inside the target nucleus. This behavior is clearly demonstrated in the comparison between our calculation and the experimental data in the kinematic region of the E866 experiment. The dominant nuclear modification of the DY cross section is from nuclear shadowing of parton distribution functions inside large nuclei as given by the EPS08 \cite{Eskola:2008ca} parameterization. The effect of medium-modified beam quark distribution caused by beam quark energy loss leads to further suppression of the DY cross section for large nuclei. However, with the quark transport parameter predetermined from the nuclear DIS experiment \cite{Wang:2009qb}, the suppression due to initial beam quark energy loss is quantitatively small. The additional suppression only becomes considerable in large $x'$ region in a large nucleus.  Since the parameterization of nPDF \cite{Eskola:2008ca} from global fitting included DY data, one should include the effect of beam parton energy loss in large $x'$ or small $x$ region.

\begin{figure}[h]
\begin{center}
\includegraphics[width=14.0cm]{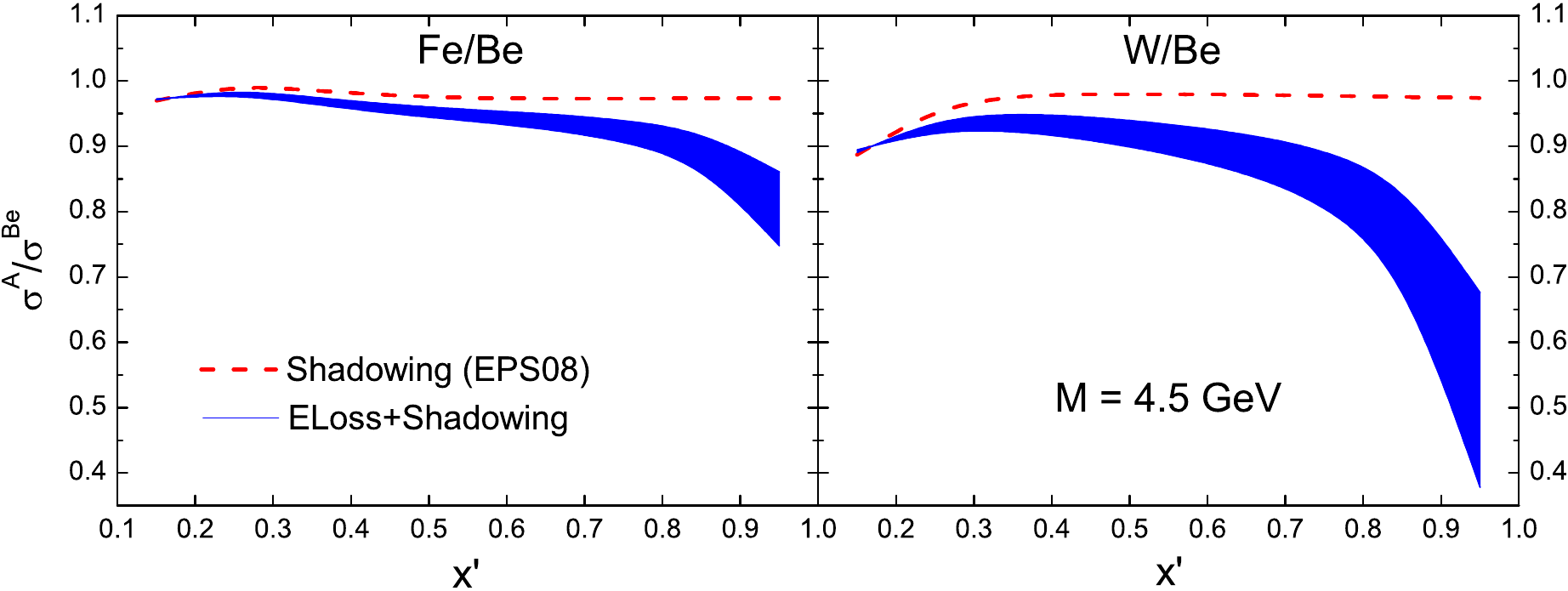}
\centerline{\parbox{15cm}{\caption{\label{Fig-e906}\small
Predictions for the DY cross section ratios per nucleon in $p+A$
collisions versus $x'$ at $E_{lab}=120$ GeV. The shaded bands correspond to
$\hat{q}_0=0.024\pm0.008~\rm GeV^2/fm$ \cite{Wang:2009qb}.}}}
\end{center}
\end{figure}
On the other hand, in p+A collisions at lower beam energy $E_{lab}$ in $p+A$ collisions, the target parton momentum fraction $x$ becomes large for moderately large beam parton momentum fraction $x^{\prime}$, where the effect of shadowing should be small.  In this kinematic region the fractional parton energy loss will become larger for smaller beam parton energy $x^{\prime}E_{lab}$ \cite{Wang:2009qb}. Therefore, at fixed DY dilepton mass $M$ and lower beam energy $E_{lab}$, one can disentangle the effect of initial-state parton energy loss from the nuclear shadowing. Shown in Fig.~\ref{Fig-e906} are the predictions for the DY cross section ratios at $E_{lab}=120$ GeV in the Fermilab's E906 experiment \cite{ref:e906web} with invariant dilepton mass $M=4.5$ GeV. At this lower beam proton energy, the effect of parton shadowing is indeed small as shown by the dashed line in Fig.~\ref{Fig-e906}. On the other hand, the energy loss effect induced by multiple scattering is significant and the dominant cause for the DY suppression shown by the shaded bands in Fig. \ref{Fig-e906}. Therefore, the E906 experiment can provide an unambiguous measurement of the effect of initial-state parton energy loss in DY cross section.

\section{Conclusion}
\label{}
We have discussed the medium modified beam quark distribution function in the DY process in p+A collisions due to multiple scattering within the framework of generalized high twist factorization formalism. The medium modification to the beam quark distribution depends on the twist-4 matrix element which can be related to jet transport parameter $\hat{q}$. Using the value of $\hat{q}$ determined from nuclear DIS data, we evaluate the nuclear
modification factor in the Drell-Yan process in p+A collisions for different nuclear targets in two kinematic regions corresponding to the Fermilab E866 and E906 experiments, respectively. We found that in E866
experiment, the effect of beam parton energy loss is only considerable in the large $x^{\prime}$ region with heavy nuclear targets. However, this effect becomes significant for lower beam proton energy as in E906 experiment.

\begin{center}
{\bf Acknowledgments}
\end{center}
This work is supported by the NSFC of China under Projects Nos. 10825523, 11205035 and by the Director, Office of Energy Research, Office of High Energy and Nuclear Physics, Divisions of Nuclear Physics, of the U.S. Department of Energy under Contract No. DE-AC02-05CH11231 and within the framework of the JET Collaboration.




\begin{thebibliography}{99}


\bibitem{Gyulassy:1990ye}
  M.~Gyulassy and M.~Plumer,
  Phys.\ Lett.\ B {\bf 243}, 432 (1990);
  X.~-N.~Wang and M.~Gyulassy,
  Phys.\ Rev.\ Lett.\  {\bf 68}, 1480 (1992).

\bibitem{Luo:1992fz}
  M.~Luo, J.~-w.~Qiu and G.~F.~Sterman,
  Phys.\ Lett.\ B {\bf 279}, 377 (1992).
M.~Luo, J.~-w.~Qiu and G.~F.~Sterman,
  Phys.\ Rev.\ D {\bf 49}, 4493 (1994).
M.~Luo, J.~-w.~Qiu and G.~F.~Sterman,
  Phys.\ Rev.\ D {\bf 50}, 1951 (1994).

\bibitem{Guo:2000nz}
  X.~-f.~Guo and X.~-N.~Wang,
  Phys.\ Rev.\ Lett.\  {\bf 85}, 3591 (2000);
  X.~-N.~Wang and X.~-f.~Guo,
  Nucl.\ Phys.\ A {\bf 696}, 788 (2001).

\bibitem{Airapetian:2007vu}
  A.~Airapetian {\it et al.}  [HERMES Collaboration],
  Nucl.\ Phys.\ B {\bf 780}, 1 (2007)
  [arXiv:0704.3270 [hep-ex]].

\bibitem{Wang:2009qb}
  W.~-t.~Deng and X.~-N.~Wang,
  Phys.\ Rev.\ C {\bf 81}, 024902 (2010)
  [arXiv:0910.3403 [hep-ph]].

\bibitem{Guo:1998rd}
  X.~-f.~Guo,
  Phys.\ Rev.\ D {\bf 58}, 114033 (1998)
  [hep-ph/9804234].

\bibitem{Kang:2011bp}
  Z.~-B.~Kang, I.~Vitev and H.~Xing,
  Phys.\ Rev.\ D {\bf 85}, 054024 (2012).
  H.~Xing, Z.~-B.~Kang, I.~Vitev and E.~Wang,
  arXiv:1206.1826 [hep-ph].

\bibitem{Xing:2011fb}
  H.~Xing, Y.~Guo, E.~Wang and X.~-N.~Wang,
  Nucl.\ Phys.\ A {\bf 879}, 77 (2012)
  [arXiv:1110.1903 [hep-ph]].

\bibitem{Vasilev:1999fa}
  M.~A.~Vasilev {\it et al.}  [FNAL E866 and NuSea Collaborations],
  Phys.\ Rev.\ Lett.\  {\bf 83}, 2304 (1999)
  [hep-ex/9906010].

\bibitem{Eskola:2008ca}
  K.~J.~Eskola, H.~Paukkunen and C.~A.~Salgado,
  JHEP {\bf 0807}, 102 (2008)
  [arXiv:0802.0139 [hep-ph]].

 \bibitem{ref:e906web}
E906 home page: http://www.phy.anl.gov/mep/SeaQuest/

  \end{thebibliography}



\end{document}